# Urban Building Energy Modeling (UBEM) Tools: A State-of-the-Art Review of bottom-up physics-based approaches

Martina Ferrando[1*], Francesco Causone[1], Tianzhen Hong[2], Yixing Chen[3,4]

[1] Department of Energy, Politecnico di Milano, via Lambruschini 4, Milan, Italy
[2] Building Technology and Urban Systems Division, Lawrence Berkeley National Laboratory, California, USA
[3] College of Civil Engineering, Hunan University, Changsha, 410081, China
[4] Key Laboratory of Building Safety and Energy Efficiency (Hunan University), Ministry of Education, China

*Corresponding author

martina.ferrando@polimi.it

Energy Department, Politecnico di Milano
Via Lambruschini 4, 20156 Milano MI

# Urban Building Energy Modeling (UBEM) Tools: A State-of-the-Art Review of bottom-up physics-based approaches

## Abstract

Regulations corroborate the importance of retrofitting existing building stocks or constructing new energy-efficient districts. There is, thus, a need for modeling tools to evaluate energy scenarios to better manage and design cities, and numerous methodologies and tools have been developed. Among them, Urban Building Energy Modelling (UBEM) tools allow the energy simulation of buildings at large scales. Choosing an appropriate UBEM tool, balancing the level of complexity, accuracy, usability, and computing needs, remains a challenge for users. The review focuses on the main bottom-up physics-based UBEM tools, comparing them from a user-oriented perspective. Five categories are used: (i) the required inputs, (ii) the reported outputs, (iii) the exploited workflow, (iv) the applicability of each tool, and (v) the potential users. Moreover, a critical discussion is proposed, focusing on interests and trends in research and development. The results highlighted major differences between UBEM tools that must be considered to choose the proper one for an application. Barriers of adoption of UBEM tools include the needs of a standardized ontology, a common three-dimensional city model, a standard procedure to collect data, and a standard set of test cases. This feeds into future development of UBEM tools to support cities' sustainability goals.

## 1. Introduction

The 2030 agenda for sustainable development, signed by all 193 governments of the United Nations, includes 17 Sustainable Development Goals (SDGs). Between them, the SDG 11 "Sustainable cities and communities" [1], stresses the importance of cities and settlements to improve living standards and decrease energy consumption. In particular, buildings are one of the main contributors to energy and materials used in urban areas all over the world [2,3]. In this scenario, energy in cities must be better managed, and the cities' design must be optimized [4]. Thus, numerous urban building energy modeling (UBEM) methodologies and tools have been developed [5]. They provide users with the energy demand of the building stock, comprehending benchmarking analyses, scenarios evaluation, peak loads and energy patterns analysis, and other peculiar analyses [6]. UBEM is different from the other two approaches used to model the urban environment: land-use/transport analyses (reference tools are UrbanSim [7–9], OPUS [10], ILUTE [11] and ALBATROSS [12]) and urban system energy modeling (USEM). USEM is specifically focused on optimization and design of energy networks and systems [13], usually modeling the building stock without a high level of detail (reference tools are HOMER [14], EnerGIS [15], Syncity [16,17], iTEAM [11], reMAC [18] and PVGIS tool [19]). Conversely, the aim of UBEM is primarily the modeling of the building stock, that can be used eventually as input to properly design and optimize energy systems; thus, in the latest tools, UBEM and USEM approaches are brought together. *Figure 1* represents a schematic of urban modeling approaches, in which, in light grey the subcategory on which this paper is focused is highlighted.

In recent years, the topic of urban building energy modeling has been extensively reviewed. One of the most acknowledged reviews in the field of urban modeling of the residential sector, in terms of classification and methodologies, was by Swan and Ugursal [6]. They started subdividing the modeling methodologies in top-down and bottom-up approaches and then, have deepened the latter for the modeling approaches. The focus on the bottom-up approach has been enhanced with the review work of Kavgic et al. [20], in which the aims of these models are also addressed. In the last two years, the bottom-up approach has gained momentum and new reviews focused on more specific aspects have been published. The work of Reinhart and Cerezo Davila [4] is focused on the subtasks that a modeler should undertake to model urban energy demand; they have divided the review into three steps: data input, thermal modeling, and validation. Li et al. [21] have given an updated review of the modeling approaches and procedures available to model urban building energy use. Cerezo Davila et al. [22], in the first part of their work, have given a review of the different inputs needed to create urban building energy models. While, Nageler et al. [5] comparing two different dynamic approaches, have presented a good overview of possible methods in modeling urban environments. An updated broad and general overview of modeling energy demand at the city scale is given in the works by Frayssinet et al. [23] and Ferrari et al. [24,25]. The review of Mauree et al. [26] gives a comprehensive summary of the assessment methods for sustainability at an urban scale, including tools for analyses of microclimate and comfort. The novel work of Hong et al. [27], presented in a form of ten questions on UBEM, covers all the essential steps of the modeling process and the next development potentials of UBEM research and applications. Other reviews are focused on related specific topics, such as the work of Naber et al. [28], centered on how uncertainties are handled in urban models, the work of Happle et al. [29], focused on occupant behavior modeling at the urban scale, or the work of Chen et al. [30], concentrated on the creation of proper city building dataset for use in UBEM.

However, a user-oriented overview, as already done for building energy modeling (BEM) programs by Crawley et al. [31], is missing, and it is, thus, the aim of this paper. Each available tool focuses on some aspects and overlooks others [32]; furthermore, the peculiarities are often not explicitly defined. UBEM is a multi-disciplinary and complex field [33], in which a compromise must be achieved between detail of representation, model accuracy, usability, data quality, and computational effort [34,35].

The UBEM approaches are usually divided into two main categories [20,21,36]: top-down and bottom-up (*Figure 1*). The top-down models do not need any specific data related to each single building, they can estimate the energy consumption of buildings from long-term relationships that link the energy used by buildings and some drivers. The nature of these drivers divides this typology of models in three main subcategories: socio-econometric [37,38], technical [39,40], and physical models [20,41]. In these models, less input data is needed, and long-term socio-economic aspects can be included. However, they cannot intrinsically predict future trends because they are based on past interconnections and, usually, they give only aggregated data as results, without any precise spatial or temporal detail. On the contrary, the bottom-up approach is based on the calculation of each single building's energy consumption, which is subsequently aggregated within an integrated framework. Based on how the energy consumptions are calculated, two different approaches are identified. The statistical, called also data-driven, models make use of data mining and machine learning techniques to evaluate the energy consumption of buildings [42,43]. Whereas, the physics-based, or engineering, models calculate the energy consumption of buildings via a detailed thermal characterization [44]. Even if these models need a vast amount of data to characterize every single building in the stock and can require significant computational effort, they are versatile enough to identify the strategic mix of policies and renovation measures for large building stocks [33]. Moreover, the other typologies of models mainly predict annual energy consumption without the proper spatiotemporal detail. The aim of the bottom-up physics-based UBEM is also to design and optimize urban energy systems and to plan urban development [29]; thus, their results are characterized by a higher spatiotemporal detail compared to other methods. The physics-based models can further breakdown into detailed multi-zone dynamic thermal simulation models and reduced-order resistor-capacitor (RC) models based on the simulation method. The first models use detailed dynamic heat balance equations of the walls, zones and energy systems; while the seconds make use of RC models in which, usually, each building is represented by a single thermal zone.

Bottom-up physics-based UBEM is a relatively new field of study. It takes its origin from BEM; however, major differences are outlined between the two approaches. The presented work is intended as a user-oriented review aiming to give a guideline to BEM modelers that are interested in urban scale simulations or to people from institutions interested in understanding which are the possibilities in this field. It gives a clear overview of which are the main available bottom-up physics-based UBEM tools and their further advancement. The tools are developed for different goals and make use of different simulation engines and Graphical User Interfaces (GUIs); thus, a comparison can help to understand the strengths and limitations of each one. Five feature categories are chosen to describe and compare the tools: (i) input, (ii) output, (iii) workflow, (iv) applicability, and (v) potential user. These categories are fundamental from a user point of view because underlying the phases that a modeler must face in UBEM (e.g., collection of data, modeling, results analyses) and are the main categories that must be taken into consideration in the choice of one tool respect to others. This paper focuses on the bottom-up physics-based UBEM tools available at the current state. The main outputs of this paper are: (i) to provide a summary description of the already developed and the promising tools (*Section 2*), (ii) to compare the main tools (*Section 3*), and, (iii) to point out the research development potentials on which the different teams are working on at the current state (*Section 4*). While *Section 3* is specifically focused on a comparative analysis of the tools, *Section 4* summarizes the main advancements that are currently under study for the future development of these tools. This will inform users and stakeholders on how to choose an appropriate UBEM tool for a specific application, as well as UBEM developers and researchers on future opportunities and improvements of UBEM tools for wide adoption.

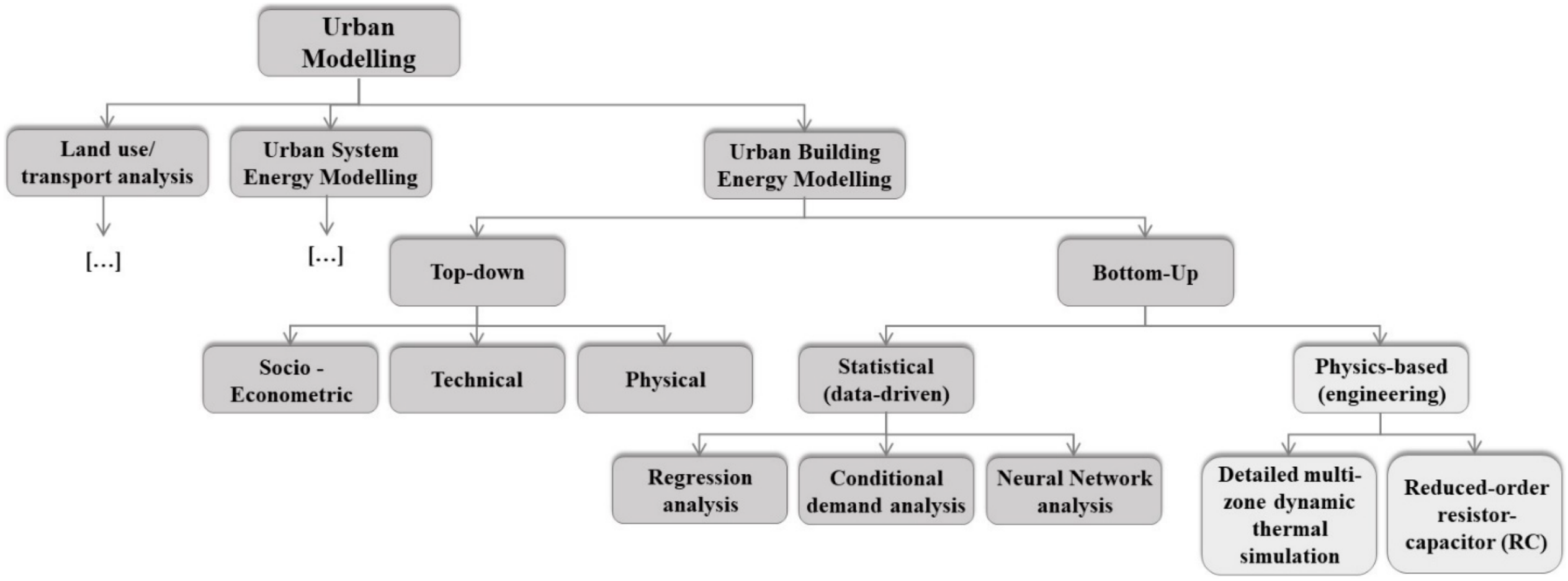

*Figure 1: Schematic of the Urban modeling approaches, in light grey the subcategory analyzed in this paper*

## 2. Bottom-up physics-based UBEM tools

Numerous methodologies for UBEM have been developed during recent years, however, they are strictly correlated to specific case studies and usually, they do not have dedicated GUIs [45–50]. In this review, only the bottom-up physics-based tools created expressly for urban applications and that have been used for different case studies have been included. They allow modeling in detail the building stock with a physics-based approach.

In 2009, starting from the Sustainable Urban Neighbourhood modeling tool (SUNtool [51]), CitySim [34] was developed in Java and C++. It aims to support the sustainable planning of urban settlements and can simulate the energy use of a few buildings up to tens of thousands. The thermal model of CitySim is based on an equivalent electrical circuit that allows considering also subspaces in buildings, linking them through separating walls' conductance. In 2013, SimStadt [52] was developed as an urban energy simulation platform to support the planning of the energy transition at the urban scale. It allows the fast creation of evaluation scenarios, using refurbishment rates (percentage of floor area per year), time horizon and priority indexes. It is developed as a Javascript and it is integrated with the three dimensional (3D) city model format CityGML [53] and Energy ADE [54] (i.e., an extension for CityGML to describe the buildings' fabric and technical systems). In the same year, the urban modeling interface (umi) [55] was developed aiming to evaluate the energy use of buildings at neighborhood and city-scale, sustainable transportation choices, daylighting, outdoor comfort and food production [56,57]. The tool uses Rhinoceros [58] as a Computer-Aided Design (CAD) modeling. It is integrated with the Urban Weather Generator tool, accounting urban weather effects and Daysim [59], for daylight availability. The years 2015 and 2016 saw a strong development of bottom-up physics-based tools. In 2015, City Building Energy Saver (CityBES) [60] was released as a web-based platform able to simulate the energy performance of buildings at large-scale [33]. The main intended use cases are: energy benchmarking, urban energy planning (to evaluate the best strategies to optimize the energy systems), energy retrofit analysis (to evaluate different scenarios of retrofit and to support decision-makers) and building operational management (to improve operations of the building stock at urban scale), evaluation of solar PV potential, and visualization of urban microclimate. CityBES can interface CityGML [53], with the characterization of the building stock and the simulation engine to allow the visualization of buildings' performance. In the same year, Open Integrated District Energy Assessment by Simulation (OpenIDEAS) was developed as an open-source Modelica-based framework. It allows the assessment of building load profiles for analysis of optimized district energy systems. It can simulate buildings integrated with an energy network at the district level with a focus on low-voltage grids. It allows the integration of statistical methods for schedule and input samples too [61,62]. While, in 2016, City Energy Analyst (CEA) [63] was released, based on Python [64]. It is characterized by a friendly GUI that helps in the management of data. It is able to run energy simulations and directly compare scenarios. In the same year, Urban Renewable Building and Neighborhood optimization (URBANopt) was developed as an application to simulate the energy performance of low energy districts, including the option for district heating and cooling systems [65]. It uses the OpenStudio platform to perform detailed energy modeling at the individual building level leaning on EnergyPlus. The simulation platform adopts NREL's Customer Optimization For Furthering Energy Efficiency (COFFEE) [66], which is an application to generates baseline energy models of buildings. Finally, in 2018, Tool for Energy Analysis and Simulation for Efficient Retrofit (TEASER) [67] was released based on Python [64]. It aims to integrate the UBEM with USEM, thus, it is able to describe in detail the built environment at the city scale allowing the deep characterization of urban energy systems including distribution. It was developed for the fast assessment of energy efficiency potentials of the building stock, combining multiple datasets to characterize buildings and running dynamic simulations to assess their energy use. In *Table 1*, a summary of these tools is reported, and in *Figure 2* a timeline of the presented bottom-up physics-based UBEM tools is presented.

*Table 1: Main characteristics of the selected tools*

| Tool | CitySim | SimStadt | umi | CityBES | OpenIDEAS | CEA | URBANopt | TEASER |
|---|---|---|---|---|---|---|---|---|
| Year | 2009 | 2013 | 2013 | 2015 | 2015 | 2016 | 2016 | 2018 |
| Developer | EPFL | University of Stuttgart | MIT | LBNL | KU Leuven | ETH Zürich and Singapore | NREL | RWTH Aachen University |
| URL | https://citysim.epfl.ch/ | http://www.simstadt.eu/de/index.jsp | http://web.mit.edu/sustainabledesignlab/projects/umi/ | https://citybes.lbl.gov/ | https://github.com/open-ideas | https://cityenergyanalyst.com/ | https://www.nrel.gov/buildings/urbanopt.html | https://github.com/RWTH-EBC/TEASER |
| Status | Actively maintained | Active, under development | Active, under development | Active, under development | Actively maintained | Active, under development | Active, under development | Active, under development |
| Availability | Free | Not publicly released | Free, but needs Rhinoceros 6 license | Free, via developers' support | Free | Free | Not publicly released | Free |
| Modeling approach | Reduced-order RC Model (CitySim solver) | Reduced-order RC Model (ISO 13790) | Heat-Balanced Physics Model (EnergyPlus) | Heat-Balanced Physics Model (EnergyPlus + OpenStudio) | Reduced-order RC Model (FastBuildings) | Reduced-order RC Model (ISO 13790 adapted) | Heat-Balanced Physics Model (EnergyPlus) | Reduced-order RC Model |
| Computing platform | Personal Computer | Personal Computer | Personal Computer | Web-based | Personal Computer | Personal Computer | High-Performance Computer | Personal Computer |
| Usage examples | EPFL University campus [68,69] , La Jonction" district (Geneva) [70], Nablus [71] | Grünbühl (Ludwigsburg), Rintheim (Karlsruhe), Bospolder (Rotterdam) [72] | Boston [22], Kuwaiti city neighborhood [73] MIT University campus [74] | Different districts of San Francisco [75–77] | Belgian city block [61], Belgian residential zero-energy neighborhood [62] | One central area in Zurich [78], City block in Zug [79] | National Western Center and Sun Valley district [65] | Research campus in Germany [67], Bad Godesberg (Bonn) [67] |
| Main Structure | Four core models: (i) thermal model, (ii) radiation model, (iii) behavioral model, (iv) plant and equipment model + possible integration with Multi-Agent Transport Simulation toolkit (MATSim-T [80]) | Energy analysis run exploiting ISO 13790 and estimation of solar potential through online database (i.e. PVGIS [19]) or weather data files (i.e. Meteonorm [81]). | Six modules: (i) daylight module, (ii) energy module, (iii) lifecycle module, (iv) mobility, (v) site module, (vi) harvest module for food production. | Three main sections (District Building, Modeling and Analysis and Urban Climate). The Modeling and Analysis section includes five sub-tools (benchmarking, retrofit scenarios, renewables, Life Cycle GHG and simulation). | Three components: (i) systems (Modelica IDEAS library), (ii) stochastic residential occupancy behavior (Python StROBe), (iii) building modeling (Modelica FastBuildings library + GreyBox) | Seven databases (weather, urban environment, energy services, conversion, distribution, systems, and targets) and six modules (demand, resource potential, system technology, supply system, decision, analysis) | Three main components: (i) 3D building energy models from map imagery and billing data, (ii) COFFEE as simulation platform and (iii) Building Component Library (BCL) for energy conservation measures and retrofit analysis | Three main packages: (i) the data package (that allows the input of data and the reading of outputs), (ii) the logic package (that helps in the manipulation of data), and (iii) the GUI package. |

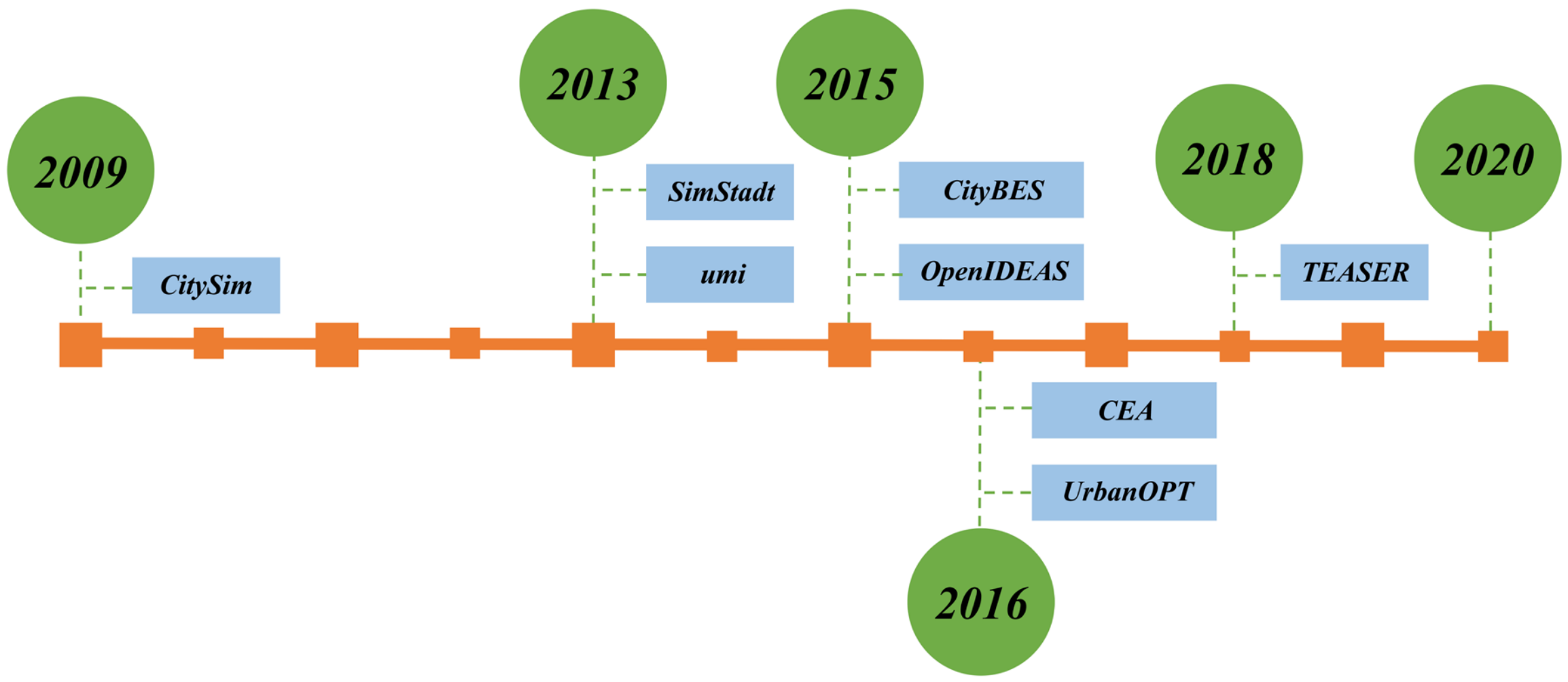

*Figure 2: Development timeline of bottom-up physics-based UBEM tools*

Numerous other tools are currently under development. An example is the System Master PLanner / Net Zero Planner (SMPL/NZP) Tool [82] developed by the U.S. Army Engineer Research and Development with the U.S. Army Construction Engineering Research Laboratory. This is a web-based tool supporting energy, water, and waste planning, estimating current and future energy loads, optimizing the supply systems to meet loads of both electric and thermal energy. It is also able to estimate costs and returns of different strategies, running return on investment analyses. The building energy calculation method relies on EnergyPlus. This tool is under development and even if it has been used numerous times by the U.S. Army, it is not yet available to the public. Other tools have been used only once, for specific case studies for which they were optimized. However, some of them used promising methods that could be applied to other case studies. For example, Energy Atlas Berlin [83], developed by the Technische Universität München, is a holistic tool for the assessment of building energy demand at urban scale and for urban energy networks, with a focus on renewable energy sources (i.e., photovoltaic and geothermal). It allows the comparison of different scenarios to optimize intelligent urban management and energy infrastructure design. The tool is based on a CityGML model for the city of Berlin and includes also transport, water, and waste management. LakeSIM [84], developed in collaboration with the University of Chicago, tries to bring together the urban planner tools with high-detail models for building energy and transportation. The aim is to achieve high-performance models to produce long-term and high-resolution estimates for energy, transportation, wastewater, and other critical issues. The test case is the 600-acre Chicago Lakeside Development on the South Side. Moreover, the Temperature of Urban Facets Indoor-Outdoor Building Energy Simulator (TUF-IOBES) [85], developed by the University of California San Diego, is a building-to-canopy model that simulates indoor and outdoor building surface temperatures and heat fluxes in an urban area to estimate cooling/heating loads and energy use in buildings via transient heat transfer balance. The heat balance considers real weather conditions, indoor heat sources, building, and urban material properties, building envelope, and systems. Building Energy Modeling - Town Energy Balance (BEM-TEM) [86] is an urban canopy model able to assess the energy effects of buildings in an urban climate and calculate the building energy consumption. It calculates the energy consumption of buildings applying a heat balance method and it was used and validated for a district of Haussmannian buildings in Paris. Finally, Virtual Electric Power Board (Virtual EPB) [87], developed by Oak Ridge National Laboratory, uses the Chattanooga EPB's 15-minute electricity consumption dataset along with Geographic Information System (GIS) data to create a calibrated digital model to support analysis regarding energy-efficiency design, demand response design, design for resiliency, and customer communications.

## 3. Tools comparison

In this section, five feature categories are used to compare the eight selected tools previously introduced: (i) the input needed to run a simulation, (ii) the output that is directly provided from simulations, (iii) the workflow exploited to run the analyses, (iv) the applicability of each tool regarding the scale and the type of project of the case study, and (v) the potential user of each tool. The comparison has been conducted at the beginning of 2020, taking information from available official documents.

## 3.1. Input

*Table 2* provides a summary of the inputs required by each tool to properly run simulations. Besides the location and the corresponding weather dataset, two main steps establish the input setting phase for all the analyzed tools. Firstly, the building stock geometry must be imported and secondly, thermophysical properties must be associated with all entities in the geometry.

The setting of the geometry is the step that most diverges from the BEM tools. Single-building models can be high-detailed, also considering internal subdivisions and rooms. However, dealing with UBEM, a simplification of the geometry is required to decrease the computational time of numerous buildings together. Almost all the tools are directly integrated with GIS format files (e.g., CityGML, GeoJSON, Shapefile, OpenStreetMap). GIS formats are indeed the most common and standard method to store buildings information by the municipalities. Moreover, the CityGML format allows the setting of 3D building geometry and the assessment of its properties, including also the description of the urban environment (e.g., terrain characteristics, water bodies, vegetation, transport network, urban furniture, etc.). Umi, at this state of the art, is not integrated with these typologies of formats, but a new version allowing the integration with GIS formats is currently under development. For all tools, usually, the buildings are simplified to extrusion of the footprint for the height of the building (corresponding to the first level of detail (LOD) of CityGML files [88]) constituting a single thermal zone. When available, SimStadt, CityBES, and TEASER allow the use of CityGML's files LOD 2, thus a few more geometrical details are considered (e.g., loggias, saddle roofs). For these tools, the integration with LOD 3 (that includes windows and other openings, that are not considered in lower LOD) is under development. However, the CityGML files are not always available for urban areas and, of course, for new settlements. For this reason, manually created GIS files can be always used as input files for the geometry.

The second fundamental step is the association of the thermophysical properties to the entry geometry. The characteristics of the building fabric, of the systems and schedules of usage, must be assigned to each thermal zone of the geometry. While, in BEM, the single building is deeply characterized, in UBEM, this step is usually simplified by the use of archetypes. Archetypes are fully characterized prototype buildings, assessed to be representative of the modeled stock, exploited to quickly describe the buildings in the geometry. Usually, a large amount of data is required to derive these prototype buildings, thus, almost all the tools suggest already implemented default archetypes. The archetypes are assigned via basic and usually available data such as the construction year, the intended use (e.g., residential, office, commercial), and the typology of the building (tower, detached house, lined building, etc.). SimStadt and CityBES allow direct integration with Energy ADE [54], which is an extension of the CityGML files that includes some thermophysical characteristics of building components.

Part of the characterization regards the occupants' description. The interaction of occupants with the building is related to their presence and their actions, such as opening windows, switching lights, adjusting blinds, and the use of appliances. All the tools allow the description of occupants' presence and actions in a deterministic way (i.e., via predefined fixed schedules). Probabilistic models are under study to properly consider the intrinsic stochastic nature of occupants and to consider their movement between the different and numerous thermal zones of an urban model. OpenIDEAS's

StROBe module allows the stochastic residential occupancy behavior modeling, while, for almost all the other tools it is a feature under investigation.

Finally, some tools can require other input parameters for specific analyses. For example, SimStadt, CityBES, and CEA require also the energy conservation measures to be tested or optimized and the targets to achieve.

*Table 2: Summary of the inputs required by the tools*

| INPUT | | | CitySim | SimStadt | umi | CityBES | OpenIDEAS | CEA | URBANopt | TEASER |
|---|---|---|---|---|---|---|---|---|---|---|
| | Geographic Information System (.shp, .gdb, etc.) | | | X | S | | | X | | |
| | CityGML | | X | X | | X | | X | | X |
| | GeoJSON | | | | | X | | | X | |
| | Open Street Map (.osm) | | | | | | | | X | |
| | Intermediate Data Format (.idf) | | | | X | X | | | X | |
| | Modelica (.mo) | | | | | | X | | | X |
| | Python (.py) | | | | | | X | | | X |
| Building characterization | EnergyADE integration | | | X | | S | | | | X (as output) |
| Building characterization | Archetypes based on: | Intended use | X | | X | X | X | X | X | X |
| Building characterization | Archetypes based on: | Construction Year | | | | X | | X | | X |
| Building characterization | Archetypes based on: | Volume | | | | | | | | X |
| Building characterization | Archetypes based on: | Building typology | X | | | X | | X | X | |
| Building characterization | Default characteristics included | | | | X | X | X | X | X | X |
| Building characterization | Characterization includes: | Envelope | X | X | X | X | X | X | X | X |
| Building characterization | Characterization includes: | Systems | X | X | X | X | X | X | X | X |
| Building characterization | Characterization includes: | Energy use | X | X | | X | | | X | |
| Building characterization | Occupants description | Deterministic | X | X | X | X | | X | X | X |
| Building characterization | Occupants description | Stochastic | S | | S | S | X (residential) | P | | P |
| Other | Energy Conservation Measures | | | | | X | | | | |
| Other | Targets | | | X | | X | | | | |

X = feature or capability currently available, P = feature or capability partially implemented, S = feature or capability under study

## 3.2. Output

*Table 3* summarizes the outputs provided by running a typical simulation with each tool. It must be noted that, in this section, different nomenclatures are exploited by the different developers' teams to describe the outputs of their tool, hindering their comparison. Thus, general definitions are used in this section. For example, heating/cooling thermal energy is used, because it is not always possible to distinguish between energy needs and energy uses according to ISO 52000-1 [89].

The outputs can be subdivided into four main categories: (i) building-related, (ii) resource potential, (iii) urban energy systems, and (iv) large scale general evaluations.

The most developed category of outputs is the one related to building energy use, in particular, heating and cooling thermal energy, domestic hot water demand, electric use and in some cases daylight. Numerous tools provide also an estimation of the resource potential, such as the solar potential on roofs and facades but also, in the case of CEA, ambient heat potential (e.g. geothermal, lake water and source of waste heat). These results may be exploited, for example, to calculate the electric and thermal potential of the installation of photovoltaic and solar thermal panels. Other tools exploit other sub-tools to do the same (e.g., SimStadt exploiting PVGIS [19]).

The third large category of outputs is the one related to urban energy systems and the direct integration of methodologies used in USEM. For this purpose, several tools make use of already developed equation-based object-oriented district system analysis solutions (e.g., UrbanOPT exploits OpenStudio, OpenIDEAS and TEASER are based on Modelica). CEA creates a proposal for the district system geometry from OpenStreetMap and runs the simulations using a simplified network-based approach, considering a constant heat loss for the pipes and coefficients for nodes. It performs thermal network optimization and thermal & electrical grid planning (considering buildings connections, geometry, pipes dimension, and costs analysis. Moreover, it runs energy supply system optimization, minimizing annual capital costs, or annual greenhouse gas emissions or annual primary energy consumptions. CityBES uses EnergyPlus as the engine to simulate district heating and cooling systems. CityBES imports and visualizes a cooling and heating load profile of a district, then users select a few district energy system types and specify their characteristics for evaluation. As follow, EnergyPlus models are created and simulations are run to calculate the energy use and energy cost of the selected district energy systems, and finally simulation results are shown for users to compare performance between the selected district energy systems. Currently, five types of district energy systems are supported in CityBES, including water-cooled chillers and boilers, water-cooled chillers with ice-storage and boilers, heat-recovery chillers and heat pumps, and geothermal heat pump. The heat loss through the transport network is simplified as a load correction factor.

Numerous tools allow also the direct comparison of Energy Conservation Measures (ECMs) scenarios and evaluation of the resulting Greenhouse Gas (GHG) emission. A few tools show peculiar analysis (i.e., CityBES is provided with 75 already developed ECMs to be tested, CEA performs a cost-benefit analysis of the applied strategies and the new release includes electro mobility analysis, CitySim is integrated with Multi-Agent Transport Simulation toolkit (MATSim-T [80]) to perform transport analysis, whereas umi evaluates the efficiency of a district considering its walkability, bikeability, and production of food).

The typology of output is important and can hinder or promote the use of one tool respect to others. A comparison is made based on the time resolution, the spatial resolution, and the available display implemented in the tool. All the tools provide yearly results for the energy consumption of the building stock. Some tools are provided with very detailed models that reach the minute resolution (i.e., CityBES and OpenIDEAS) or the hourly resolution (i.e., CitySim, CEA, URBANopt, TEASER). Finest resolution data is eventually aggregated into coarser results in the visualization graphs. Sometimes, the time resolution of the results can differ from the time-resolution of the simulation. For example, umi runs simulations with a time step of a minute, but the results are shown as yearly aggregated values. The spatial resolution is another element of distinction among tools. In all cases, single buildings' data is directly derived from the inputs that then, can be aggregated into groups of buildings. Umi, for daylight analysis, directly shows results for a single story in each building. Besides the fact that all the tools provide outputs in the form of spreadsheet files (usually in CSV-format), they all, except OpenIDEAS and TEASER, also provide a graphic visualization of results. In some cases, the outputs are already plotted on the 3D geometry in color scales (e.g., energy consumption, $CO_2$ equivalent), in others, graphs and charts are visualized or the results are provided in GIS shapefiles that can be opened in other software for the post-processing and visualization.

*Table 3: Summary of the outputs provided by the tools*

| OUTPUT | | CitySim | SimStadt | umi | CityBES | OpenIDEAS | CEA | URBANopt | TEASER |
|---|---|---|---|---|---|---|---|---|---|
| Building-related | Daylight | X | | X | | | | X | |
| | Heating/Cooling thermal energy | X | X | X | X | X | X | X | X |
| | Domestic Hot Water demand | X | X | X | X | X | X | X | X |
| | Electric use | X | X | X | X | X | X | X | X |
| Resource potential | Solar Roofs | | X | X | X | | X | X | |
| | Solar Walls | | | X | | | X | | |
| | Other | | | | | | S | X | |
| Urban Energy Systems | District Heating/Cooling | P | | | X | X | X | X | X |
| | Electric Grid | P | | | | X | X | X | X |
| | Energy Storages | P | | | S | X | S | X | X |
| Large scale general evaluations | Scenarios | X | X | | X | | X | X | |
| | Benchmarking | | | | X | | X | | |
| | Cost-Benefit analysis | | | | X | | X | | |
| | Transport/Mobility | X (MATsim-T) | | X | | | X | | |
| | Life-Cycle analysis | X | X | X | | | X | | |
| | Food | | | X | | | | | |
| Time-resolution | Minute | | | | X | X | | | |
| | Hour | X | | | X | X | X | X | X |
| | Day | X | | | X | X | X | X | X |
| | Month | X | | | X | X | X | X | X |
| | Year | X | X | X | X | X | X | X | X |
| Spatial resolution | Single building floor | | | P (only daylight) | | | | | |
| | Single building | X | X | X | X | X | X | | X |
| | Group of buildings | X | | X | X | X | X | X | X |
| Display | Spreadsheet | X | X | X | X | X | X | X | X |
| | Graphic visualization | X | X | X | X | | X | X | |

X = feature or capability currently available, P = feature or capability partially implemented, S = feature or capability understudy

## 3.3. Workflow

The workflow represents the modeling process from the inputs to the outputs and post-process of the results. All the tools, typically, use a similar workflow and a general one is proposed in *Figure 3*. The workflow is subdivided into five main steps, (i) inputs, (ii) model, (iii) simulation, (iv) outputs, and (v) post-process. The inputs step is subdivided into four categories that are the weather dataset, the building-related data (that includes the geometry and characterization data, raw or already organized as archetypes coupled with 3D models or GIS files), the utility rates (that can be exploited for benchmarking analysis or during the calibration and validation process) and finally some data regarding the district systems if needed. In particular, the building-related data and the geometry could be already combined (e.g., in GeoJSON or CityGML files). The model step is the combination of the geometry with the data both regarding the building stock and the district systems. In the building model, each part of the geometry is characterized by its thermophysical properties and with building systems characteristics. For this purpose, when available, CityGML files coupled with Energy ADE files can be used. In the district system model, the terrain and area characteristics are coupled with data regarding the system itself. The two models (building and district systems) are then combined with the weather dataset in the simulation step. The simulation is usually carried out with the help of a second tool (e.g., EnergyPlus, CitySim solver, etc.). In the next step, the outputs of the simulation are generated, and other possible analyses are carried out. Automated or manual calibration and validation could be implemented within these three central steps. Finally, the post-process step regards the manipulation of the results to be downloaded as spreadsheet files or to be visualized as graphs or on the geometry (e.g., via color scales plotted on the 3D geometry).

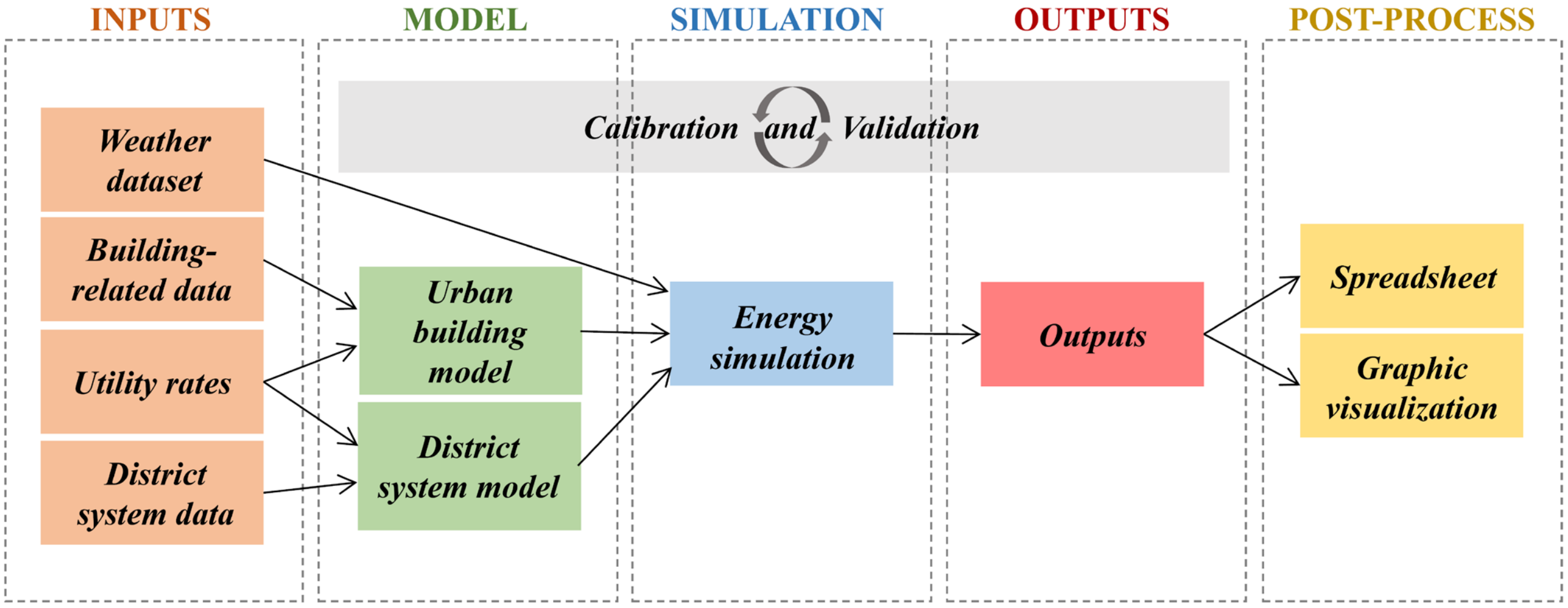

*Figure 3: General workflow used in UBEM tools*

The tool itself usually manages the model, simulation and output steps entirely (*Table 4*). Even if the simulation step, frequently, is run by an external simulation engine (e.g., EnergyPlus), the user can keep working with the UBEM tool that will manage the interoperability with the simulation engine. In most cases (e.g., for CitySim, SimStadt, CityBES, OpenIDEAS, and URBANopt) the managing of the input data is completely developed with other software and methodologies. Some of them are developed as supporting instruments to UBEM (e.g., shoeboxer [90]), whereas others are fully developed tools managing the urban geometry and its characterization allowing also advanced data analysis (e.g., cityzenith [91]). While, for umi, CEA and TEASER the GUI helps also with data management. In these tools, part of the GUI is precisely addressed to the managing of the inputs data. For example, CEA can create the building geometry and a hypothesis of district systems ducts exploiting Open Street Map. While, the only mandatory input for umi and TEASER is the building geometry, using the default settings and the GUI, the characterization can eventually be managed with the tool itself. For OpenIDEAS and TEASER, the post-processing of the results is managed separately. They are both developed directly in Python, which is a programming language often exploited to read spreadsheet files, thus, promoting the visualization and post-processing of data.

*Table 4: Workflow steps covered by the tool itself*

| STEP | CitySim | SimStadt | umi | CityBES | OpenIDEAS | CEA | URBANopt | TEASER |
|---|---|---|---|---|---|---|---|---|
| INPUT | | | X (BD) | | | X (BD and DSD) | | X (BD) |
| MODEL | X | X | X | X | X | X | X | X |
| SIMULATION | X (CitySim solver) | X | X (EnergyPlus) | X (EnergyPlus) | X (FastBuildings) | X | X (EnergyPlus) | X |
| OUTPUT | X | X | X | X | X | X | X | X |
| POST-PROCESS | X | X | X | X | | X | X | |

BD = Building-related Data, DSD = District System Data

## 3.4. Applicability

According to the previous sections, the tool has to be chosen in agreement with the available input data, the required outputs and the workflow steps that are expected to be managed directly by the tool. However, two further aspects must be taken into account in the choice of a tool: (i) the scale (i.e., number of buildings in the simulation) and the (ii) type of project (i.e., retrofit or new construction) easily managed by a tool.

The actual scale at which a tool can run simulations depends mainly on its integration with GIS formats, its simulation engine and the computing platform used. A large group of tools (i.e., CitySim, SimStadt, OpenIDEAS, CEA and TEASER) exploits the simplification derived by reducing order RC models to allow the users to run large-scale simulations directly with their own Personal Computer. Among them, CitySim, SimStadt and CEA are integrated with CityGML formats that can help in the creation of very large models. Whilst, another category of tools is the one in which multi-zone dynamic thermal simulation models are run on the hosting server or High-Performance Computer. CityBES is a web-based platform, while URBANopt makes use of High-Performance Computer to run the simulation. For this reason, a user can run a very large-scale analysis with the help of the developers' teams also due to the great integration of these tools with GIS formats file. Finally, umi runs a multi-zone dynamic thermal simulation Model (via EnergyPlus also in this case) on the Personal Computer of the user. Moreover, it is not yet integrated with GIS formats, and for these reasons, this tool is suggested to be used for medium-scale analysis.

Regarding the typology of projects, all tools can be used to run simulations on existing areas of urban settlements. In these cases, the aim of the analyses could be the increase of building efficiency, the design of urban energy systems or the evaluation of different energy conservation measures. However, when dealing with new settlements, the geometry could be a changing parameter too, thus, the integration with 3D software could be helpful. In particular, umi, being a plug-in for Rhinoceros, works very well for new settlements, also considering the possible coupling with Grasshopper and its extensions [92]. CEA's team, being the tool based on Phyton as Grasshopper, is developing new features called Grasshopper-to-CEA (GTC) [93], for the same purposes.

### 3.5.Potential Users

The user of a tool can be intended in two ways: as the modeler or as the viewer (and "consumer") of the results. The modeler is usually an expert in the BEM field and is the one who deals with the different phases of the modeling. The viewer instead wants to have a clear answer to questions and wants to extrapolate useful information from the carried analyses. Of course, these two roles are often joint into one.

The modeler, for all the tools, must have a solid experience in BEM and its main tools to deal with the setting of the simulation and its results. In UBEM, dealing with large databases and numerous information, the GUI is usually exploited to support the modeling process and the understanding of the results. Also, the developers' teams of URBANopt and CityBES give direct and complete support to use the tools and model new case studies. The GUI of umi is integrated as a plug-in for Rhinoceros, thus a good knowledge of Rhinoceros is required. While a good knowledge of Python and Modelica is fundamental to use OpenIDEAS and TEASER. For the tools that are developed in association with USEM, good knowledge of urban energy systems is required to exploit the tools at their full potential. Conversely, the viewer of the results chooses one tool respect to another for the specific output he/she is interested in. In this regard, different viewers (e.g., decision-makers, urban and building designers, district heating and cooling managers, politicians, distribution and transmission operators, modelers and researchers) could have different needs. OpenIDEAS, CEA, URBANopt, and TEASER are the tools mostly integrated with USEM, thus, they could be exploited by designers and managers of systems and by distribution and transmission operators. Umi is suitable for medium size areas and shows a more holistic point of view, thus, it can be used by users interested in different aspects at the neighborhood level (energy needs of buildings, daylight analysis, solar potential, walkability, and food production). Moreover, its results could be used by municipalities to optimize new and existing urban areas. Policymakers, modelers, researchers, urban and building designers that are interested in comparing different ECMs, can directly use CitySim, SimStadt, CityBES, CEA, URBANopt or work on different inputs scenarios in OpenIDEAS, umi or TEASER. Especially, CityBES is directly integrated with numerous ECMs and allows automatic comparisons between scenarios. For politicians that are interested in transportation together with buildings-related evaluations, CitySim integrated with MATSim-T allows this type of analysis. Policymakers interested in reducing GHG emissions can use SimStadt, CityBES, CEA or umi.

## 4. Research and development potentials

The UBEM topic is gaining momentum; tools, such as CEA, umi, and CityBES are evolving rapidly, introducing continuously new features and possibilities. Thus, the developers' teams are currently working on different challenges regarding UBEM. The critical discussion of the tools is here proposed based on the main topics under research (i.e., development of datasets for the description of buildings, the representation of people movements and actions in UBEM, the integration of UBEM with USEM aspects, the modeling of the urban microclimate, the modeling of green and blue infrastructures in cities, and the integration with the analysis of outdoor comfort, the assessment of the heat interactions between buildings and with the external environment, the calibration and validation of the UBEM models, and finally, the integration of life-cycle analysis at the urban scale). Each of these topics is deepened in a related section, focusing on the main methodologies for application to large-scale energy models and defining the advancement of the tools on the topic.

### 4.1. Datasets and description of buildings

Data in general and, in particular, datasets are essential for setting, modeling and simulating an urban energy model. Ideally, the 3D model of the urban area under study should be available and the essential thermophysical characteristics (regarding building fabric, systems, users' behavior, surroundings, etc.) should be correctly reported and assigned to the geometry. However, usually, cities do not organically collect data and a mixture from multi datasets (affected by issues related to different terminologies, and different time and spatial resolution) must be used to set the simulation. The issues of the description of buildings and the surroundings are simplified by the integration with the GIS format. Specifically, the CityGML and GeoJSON formats, allow assigning a table of characteristics (height, construction year, intended use, etc.) to a georeferenced geometry. Moreover, a detailed description of each building usually is not available. Thus, to describe buildings, the considered tools make use of archetypes, that are representative of the building stock, to quickly run simulations [94]. The archetypes are assigned usually based on building typology, construction year and other simple characteristics [95]. However, the main limitation of this approach is that the differences that are present in the real city are lost. In UBEM, this is not an issue when the results are aggregated spatially and temporally. On the contrary, when the objective is to model the urban energy systems, the spatial/temporal description should be considered. More detailed archetypes, regarding the building fabric and technical systems, could be integrated with spatial diversities within the subareas of the city. Dynamic archetypes that evolve in time, decreasing the efficiency of the systems and the building fabric, could be implemented when the time span of the simulation is long. Moreover, the diversity in terms of occupant behavior among buildings of the same occupancy type should be accounted to obtain realistic energy demand patterns and the impacts of behavioral changes over time on building energy demand. Besides the building-related data, for some peculiar analysis, also the description of the rest of the urban environment should be included (e.g., district systems, urban weather, water bodies, parks, etc.).

Almost all the considered tools can take datasets represented in CityGML or GeoJSON files (or other simpler GIS formats). In particular, umi is currently studying a direct integration with such formats. The tools currently under development are optimizing the assignment and set of archetypes, trying to include also dynamic and stochastic aspects. Moreover, a huge effort should be carried out by municipalities, to support an integrated and organic way to store and register data.

### 4.2. People movements and actions in buildings and within the city

The topic of occupant behavior (OB) modeling is well developed for BEM [96,97]. The interaction of occupants with the building is related to their presence, movement, and actions, such as opening windows, switching lights, adjusting blinds, and the use of appliances. Usually, in UBEM each typology of buildings is modeled via an archetype based on the year of construction and its intended use [94]. These archetypes include the characteristics of the envelope, systems and the schedules of typical OB. This brings no differentiation between buildings with the same intended use, resulting in non-realistic energy demand patterns [29]. However, in recent years, a few preliminary studies [98,99] pointed out that, when peak loads are important, probabilistic, space-based and person-based models must be preferred to deterministic and average value models to have a higher temporal resolution. Space-based approaches model the impact of OB aggregated into a specific space of the model (such as a specific thermal zone). The input OB parameters of person-based approaches are e.g., the aggregated power consumption of appliances used or the number of people in that space. Person-based approaches model every individual movement and actions. The individuals are modeled via archetypes (e.g., full-time employer, part-time employer, student, retiree, etc.), characterized by the presence and the control actions related to a certain space. In the context of UBEM person-based and space-based OB models could help to achieve the proper spatiotemporal resolution that is needed to have reliable and robust energy demand patterns.

All the tools' development teams manifest a large interest in this topic and numerous tools are studying or have partially integrated stochastic OB models besides the deterministic ones. In particular, OpenIDEAS's StROBe already allows the stochastic modeling of occupants in residential buildings; it is a promising technique that could be expanded also to other building uses. CEA implements a probabilistic OB model allowing different hourly schedules for the same intended use. The tools should also consider demographic variations of the population and their behavioral changes over time, moreover, in advanced models, adaptation to economic and environmental changes could be included. To achieve these advanced OB models, a huge amount of data is still needed.

Even if numerous researches deal with the modeling of occupants' actions and movements in single-buildings [100–103], the movement of people between buildings and in the city space is a fundamental and complex aspect of UBEM. Moreover, due to the high cost of sensors installation, it is unfeasible to have a dedicated sensors-based database regarding the movement of people in a city. Thus, due to the advancement of the Internet of things, numerous researchers begin to exploit the data from mobile phones, and Wi-Fi sensors to estimate the spatio-temporal movement patterns of occupants in the building and within the city [104–106]. However, proper integration of actions and movements of citizens in urban contexts is still missing, as well as a direct integration in the UBEM tools. In particular, an integration between UBEM tools and more focused urban transport network modeling and tools could be beneficial for mainly three aspects. (i) People move in the city through transportation, thus occupancy schedules can be strongly improved in an integrated approach. Study how and when people use transportation means knowing when they will reach one building or another and when they will depart from those buildings. (ii) With the advent of electric vehicles, transportation could also represent an additional and significant load, while recharging, or storage to buildings, that cannot be neglected. (iii) Zoning optimization analyses (especially, in $CO_2$ emission reduction perspectives) should consider the movement of people via transportation.

In this regard, only CitySim through MATsim-T and umi are directly integrated with a model that allows the simulation of people's movement. Specifically, MATsim-T [107] is a multi-agent transport simulation system that allows modeling people and vehicle flows and their interactions. While the mobility module of umi runs simplified analyses based on the accessibility of amenities in a district by walk or by bicycle, based on their distance. Still a lot of work is needed for the full integration of UBEM with comprehensive mobility models. Besides MATsim-T, other promising mobility modeling tools exist, such as TransSim [108] and SUMO [109].

### 4.3. Modeling of district energy systems, energy storages, and energy network

The focus of the analyzed bottom-up physics-based UBEM tools, by definition, is on the proper thermophysical modeling of the building stock. However, in recent years, due to the increase of decentralized generation of electricity, to the advancement of multi-generation systems, and the growth of integration of renewable energy sources, the urban energy systems became more interconnected and flexible [110]. These district energy systems are usually easier to be managed and should be able to serve multiple users and make the best use of energy coming from one or more energy sources [111]. The interconnection between a district system and a group of buildings is an aspect under research because fundamental for the proper design and optimization of both buildings and systems [112]. The energy networks, energy storages, centralized and single-user renewable energy systems should be included in the simulation and of course connected to the energy demand of buildings coming from their modeling.

Urban energy systems should be able to satisfy the energy demand of an urban area managing the exchange of energy [113]. This aim could be achieved by running two different analyses, firstly with a UBEM tool, to assess the energy demand and then with a USEM tool to design the systems; however, this approach hinders the optimization on both sides (buildings and systems). USEM models do not model the buildings with high detail leading to unrealistic peak demands and energy patterns or basing the design and optimization of systems on old energy patterns. Bottom-up UBEM models can help to predict current and future energy demand with the proper spatio-temporal resolution to design and optimize these systems. Thus, researchers are trying to couple the USEM to UBEM, having detailed modeling of buildings that can be used as input for the systems' design and optimization. Some of the considered tools (in particular, CEA and TEASER, but also CityBES, OpenIDEAS, and URBANopt) are born already with this aim. CitySim is partially integrated with the modeling of district heating and cooling, energy network and energy storages and it has been coupled with energy systems models [71]. In contrast, USEM tools such as Syncity [16], iTEAM [11], and reMAC [18], are trying to increase the detail in the modeling of buildings, approaching the UBEM tools [114].

### 4.4. Modeling of city microclimate, of green and blue infrastructure in cities and their integration with outdoor comfort assessment

In building simulation, the weather dataset is accounted for as one of the largest uncertainties [115]. Moreover, urban effects on climate (e.g., heat island effect, the presence of green areas and water bodies) should be considered in the simulations [116]. Usually, the weather datasets are created by measurements in rural weather stations, commonly registered at airports, and based on decades-old data. Thus, local and updated weather datasets should be implemented. Two main approaches can be used to solve the issue: modify an already existing dataset or directly use real-time measured data to create a proper weather dataset. A rural weather dataset can be converted into an urban weather dataset via models able to exploit the characteristics of the urban area and to account for urban effects. Thus, systematic frameworks to account for all the effects on the climate have been developed. An example is the Urban Weather Generator (UWG) model [117,118], which can create an hourly urban weather file from a rural one with its four modules. The first one computes the sensible heat fluxes related to the rural weather station; the second one computes the temperatures of the air layers above the rural weather station; the third one computes the same temperatures above the urban area and the last one calculates sensible heat fluxes, air temperature and humidity in the urban area. Other examples of urban weather models are the Weather Research and Forecasting model [119], and the Urban Multi-scale Environmental Predictor [120]. Another major aspect that must be considered in the creation of local weather datasets is the accounting of vegetation and water bodies, which are also known as Green and Blue Infrastructures (GBIs). GBIs play a key role through shading, evaporative cooling, and the modification of airflows and heat exchanges [121]. In the last years, researchers have investigated the role of GBIs in urban areas, and numerous methods have been developed [122]. Nowadays, for this purpose, other supporting software, as ENVI-met [123], are exploited to run the urban microclimate simulations with a detailed GBIs model. The boundary conditions could be derived from the original weather file and by the project features also used for the generation of the urban weather. The averaged output of this simulation can be used to create a customized weather file that can be used in the energy simulation [124].

The coupling effects of buildings with the microclimate and vice versa are fundamental to set reliable weather datasets, but also to account, assess and optimize the outdoor comfort in an urban environment. Specific modeling software, such as Rayman [125], SOLWEIG [126], ENVI-met [123], or CityComfort+ [127] can be exploited to estimate the mean radiant temperature thus the main thermal comfort indexes. These are modelling tools, in which the meteorological data are used as inputs, and the urban environment (considering both the geometry of the surfaces and their characteristics) has to be modeled. The automatic integration of such software with UBEM can be beneficial for the assessment of the outdoor comfort conditions.

Regarding urban weather datasets, umi is directly integrated with the UWG model, while the other tools can use weather datasets previously created with the UWG or similar models as input parameters. CitySim has been coupled with the Canopy Interface Model exploited as an interface between the atmospheric and energy models and for the turbulent fluxes in the urban canopy layer [128,129]. The tools allow the uploading of a weather dataset in almost all the common formats (e.g., epw, txt, ddy, etc.). Almost none of the mentioned UBEM tools has direct integration with a more detailed GBIs modeling and with the outdoor comfort assessment.

Some developer teams are currently improving the heat exchange among buildings and between the buildings and the surroundings, including GBIs [70].

### 4.5. Heat exchange between buildings and with the external environment

Bottom-up physics-based UBEM approach deal with the numerical representation of the complex interconnection existing between the buildings and the surrounding environment [27]. The surrounding represents the boundary condition of the building stock simulation, thus has influences on it. In contrast, buildings have a strong impact on the surroundings (e.g., shadows, heat emission, etc.) changing the surfaces' temperatures and the urban microclimate. As a matter of fact, heat emissions come from the envelope, air exchanges across openings, and building systems. Two levels of interactions can be defined: among buildings, and between buildings and the surroundings [27].

Mostly, the interactions rely on longwave radiation exchanges [130]. They occur between the external surfaces of buildings and the sky (with low effective temperature), the ground (whose temperature usually changes slower than the one of the air), and other surfaces (including other buildings' surfaces, GBIs, etc.).

The traditional building simulation tools over-simplify these exchanges, thus, in the last years, numerous researchers tried to solve the problem. Evins et al. [131] proposed a methodology and an extended review regarding the simulation of external longwave radiation exchange for buildings. Luo et al. [132] implemented in EnergyPlus version 8.8 a new feature to consider the longwave radiation between the exterior surfaces of buildings, that could be adopted in UBEM tools.

All the tools should take into account shadows while running simulations; however, nothing official is reported by the teams of SimStadt, OpenIDEAS, and TEASER. CitySim and EnergyPlus (thus, the tools relying on it) can consider shadows. CEA allows shadow analysis, making use of Radiance.
It should be noted that the interaction between buildings and the surroundings is mainly simplified. At the current state, CitySim can consider longwave radiation exchanges between exterior surfaces. Moreover, Luo et al. [132] presented a new EnergyPlus feature able to calculate the heat emissions from buildings towards the ambient air. This work could have a great impact on all the UBEM tools that rely on EnergyPlus as a simulation engine.

### 4.6. Calibration and validation of models

Models are just a representation of the real world [133]; nevertheless, they should be able to describe the meaningful characteristics of real objects. In the case of a city or a district, the modeling process aims to simplify in a structured way the reality to mirror the city through computational representation [134]. This mixture of simplification (e.g., building characterization, energy networks representation, etc.) brings to inevitable uncertainties that are difficult to be managed and quantified. According to Naber et al. [28], uncertainties permeate all steps of UBEM's workflow. All building energy models (i.e., BEM and UBEM) need to be calibrated and validated to increase the reliability of the results. Numerous manual and automated methods to manage calibration in BEM are available [135]; however, they are inappropriate for UBEM applications due to the high number of simulations involved or the large amount of data needed. In UBEM, the computation effort is a strong limitation and running numerous simulations for each building is impracticable [136]. Moreover, UBEM suffers from the typical barriers of data availability and data granularity [28]. These issues are exacerbated at such a large scale because comprehensive and detailed building stock and census data are needed. Moreover, these data are usually a combination of different sources which introduces additional uncertainties into the models due to different nomenclatures or formats. Due to this lack of reliable data, often only aggregated values are available and useable for the model validation [137]. However, this is not a true calibration process and brings to a minimal agreement between the single building measurements and the simulation results. This is due to the characterization process, typically used in UBEM, that involves archetypes. The applications of Bayesian calibration methods [22,137–140] is seen as a valuable solution in the UBEM context.

The Bayesian calibration method differs from other conventional methods because it implements, changing the uncertainty distributions, an iterative process on the inputs to better fit the output to truthful data. Thus, the final aim of the method is not to simply minimize the difference between outputs and truthful data but the tool works on the likelihood to end out with a result on the base of the uncertainties of the input parameters. However, Bayesian calibration typically requires several thousands of simulations for each building to narrow down the uncertainties.

To reduce the number of simulations per building using Bayesian calibration, Nagpal et al. [136] proposed a methodology for auto-calibrating UBEM using surrogate modeling and data-driven approximation techniques. It uses latin hypercube sampling to generate a few hundred samples and run EnergyPlus simulations for those samples to create the training datasets. Different algorithms are used to create mathematical models of the physical behavior of building systems based only on the training data. Once trained using around 200 generated training samples, the surrogate models provide instant feedback associated with changing individual building characteristics. This method vastly reduces the computation time required by optimization routines.

Chen and Hong [141] introduced an automatic and rapid calibration of UBEM based on the annual site and source energy use by learning the correlations between key model input parameters and the building energy use from USDOE reference buildings. A case study was presented to calibrate 112 large office buildings built before 1978 in San Francisco. It performed 1000 EnergyPlus simulation of the reference building to create an energy performance database. And it took another 2.7 simulation runs on average to calibrate each building rapidly. However, the methods are only tested for large office buildings in San Francisco. The applicability of the method for other climate zones and other building types is under study.

The tools' developer teams are currently working on the topic, especially for the more active tools (i.e., umi, CityBES, CEA, URBANopt). In particular, the Bayesian method is seen by most teams as a good approach to deal with the calibration. For example, CEA's team is working on a Bayesian statistical method for calibration against registered data that determines the ranges of the probability of input parameters responsible for uncertainty in the model and soon specific documentation will be provided.

### 4.7. Life cycle assessment analysis at the urban scale

The decrease of operational energy demand is not the only aim of UBEM tools, and the proposed policies and strategies should include their complete environmental impact, via Life Cycle Assessment (LCA) analysis [142]. LCA has been already applied in the UBEM context several times [143–147]. In these cases, the approach to LCA, generally, employ two steps. Firstly, the building stock is simulated, and the operational energy consumptions are assessed; secondly, the LCA is applied to the inputs and outputs and the environmental impacts are assessed. These two steps could be run in a loop for optimization analysis. LCA is typically subdivided into four main steps: (i) goal definition and scopes, (ii) inventory, (iii) impact assessment, and (iv) interpretation. In the goal definition, two main applications of LCA methods in UBEM can be identified: evaluation of the environmental impacts at the current state and in future scenarios. The main goal is the assessment of the sustainability of different energy conservation measures. The choice of the functional unit is heterogeneous (e.g., absolute, spatial or per capita), and usually, it depends on the objectives of the study. The service life widely differs too. For new buildings, a typical value is from 50 to 100 years, whilst for existing buildings a residual life is usually set, varying from 35 to 50 years. Regarding the definition of the scope, a large heterogeneity is reported both for the boundary of analysis (e.g., buildings, open spaces, energy networks, and mobility) and for the phases to be considered (e.g., production, construction, operation, and end-of-life). In the inventory step, a distinction is done between foreground and background data. The first deal with data regarding materials, construction processes, building operation, end of life and mobility. These data are usually extracted from GIS databases, registers, statistical databases, surveys, etc. While, the second is usually derived from LCA databases (e.g., Ecoinvent, Gabi), and regards extraction of raw materials for building components and transport, production of electricity, etc. For the impact assessment, the mid-point impact assessment method is usually employed. In particular, Global Warming Potential and Primary Energy are typically chosen as impact assessment indicators. Finally, the interpretation step usually regards contribution, uncertainty and sensitivity analysis and spatial visualization of the results.

Some studies (e.g., Heeren et al. [143], Anderson et al. [144], Allacker et al. [147]) propose to exploit archetypes also to run complete LCAs on representative buildings to then aggregate the results at a macro-scale. However, the current methodologies are focused on LCA without a direct integration in the bottom-up physics-based UBEM tools, mainly due to the difficulties in the proper description of buildings in such models. In the last year, a promising solution to simplify the LCA in the BEM context is through the integration with Building Information Modelling (BIM) [148,149]. Exploiting BIM for the creation of archetypes could lead to an easier integration between UBEM and LCA. Numerous examples of LCA methodologies applied to UBEM are present in the literature [150–153]. However, integrating these detailed methodologies in developed UBEM tools is not easy, because of the large number of buildings and building typologies involved. Some tools (i.e., CitySim, SimStadt, umi, CityBES, and CEA) provide the visualization of the equivalent $CO_2$ emission resulting from the implementation of some strategies; however, the framework settings of the analyses are not clear (i.e., functional unit used, service life and phases considered, inventory implemented, etc.).

## 5. Conclusions and future outlook

In this paper, a user-oriented review of the main bottom-up physics-based UBEM tools has been conducted. The comparison is exploited to define the state of the art of the topic. First, in *Section 1*, a classification has been proposed to categorize the main UBEM approaches and a focus on bottom-up physics-based UBEM tools has been made. Next, in *Section 2*, a description of the main bottom-up physics-based UBEM tool is presented. Then, in *Section 3*, a user-oriented comparison has been done based on five feature categories: (i) the input needed to run a simulation, (ii) the outputs provided by a simulation, (iii) the workflow exploited to run the analyses, (iv) the applicability of each tool regarding the scales and the type of project of the case study, and (v) the potential user of each tool. Lastly, in *Section 4*, the main challenges regarding these tools are identified and the main potential advancements are pointed out.

The bottom-up physics-based UBEM tools, dealing with the numerical representation of interconnections between the buildings and the surrounding environment, can assess the energy demand of buildings with a high spatio-temporal resolution. Thus, numerous tools have been developed in recent years. UBEM deals with the complex representation of large areas and a full-detailed simulation is not possible yet. Thus, each tool overlooks some aspects of the UBEM simulation, while, deepens some others. Potential users should consider both the intrinsic characteristics of the tool (inputs, outputs, simulation engine, etc.) and the practical limitations (integration with GIS, visualization of the results, etc.). A comparative analysis of a common case study might be implemented to review a few of the tools, to go in detail in the creation of the input files, to contrast the simulation engines and the results quantitatively. However, this cannot be done for all of the tools deepened in this paper, because their purposes and approaches showed to be substantially different, i.e., it is not easy to define a testbed working for all of the tools.

The research on the topic is growing constantly and the challenges are numerous; specially, some bottlenecks hinder the use and implementation of such tools. Among the main issues outlined from the analysis, there is the use of mix databases that impedes the fast creation of large models and the use of a dissimilar nomenclature, from one tool to another, that hampers the comparison between the tools. A lack of data needed for UBEM is registered too. A structured way to collect data should be used to promote integration within datasets and to guarantee a proper spatial distribution. At the current state, the validation process also suffers from this lack of data. Measured data should be preferred and should be integrated with data used for the characterization step. This data could also be organized in reference buildings (e.g., U.S. Department of Energy commercial reference buildings [154]) that could be exploited in both the creation of the model (in particular, for the characterization of the geometry) and its validation. A common city model and a standardization of the datasets could help in the use of the tools; a common nomenclature (for example based on the ISO 52000-1 [89]) could simplify the process of comparison and use of the tools; finally, a standard method of test (such as ASHRAE Standard 140 [155] for BEM programs) could be beneficial also for the developers' teams. Moreover, currently, UBEM tools are coupled with other software to undertake some specific analyses (e.g., evaluation of microclimate, GBIs, LCA, etc.). However, this hinders access to the common user to these analyses; in fact, the management of the input-output within different software is usually problematic. In the future, the available computational power will rise, and, by consequence, a complete integration between different tools, exploiting the GUI of the UBEM tool, could be carried out. Especially, the heat exchange among buildings and between the buildings and the surroundings could largely improve the simulation of the building energy use itself but also the analysis regarding heat island effect and outdoor comfort conditions, both within major trend research topics [26,156]. Finally, two major trends of development for UBEM tools can be defined from the present analysis. Probably, some of the tools will be optimized to run simulations at the city scale, thus, allowing the comparison of policies at large scale, with a lower detail in the characterization and huge exploitation of web-based calculation. Other tools seem, conversely, to be focusing more on the district scale, with higher attention to details, and allowing the optimization of building and district systems and the management after construction. This second group of tools will be probably promoting a major integration with BIM or similar building descriptions.

Findings from this review inform users and stakeholders on how to choose an appropriate UBEM tool for a specific application, as well as UBEM developers and researchers on future opportunities and improvements of UBEM tools for wide adoption.

## Acknowledgments

Ferrando and Causone's work was supported by the European Union's Horizon 2020 research and innovation programme under grant agreement No. 691895 - project SHAR-LLM (Sharing Cities). Hong's work was supported by the Assistant Secretary for Energy Efficiency and Renewable Energy of the United States Department of Energy under Contract No. DE-AC02-05CH11231. Chen's work was supported by the National Natural Science Foundation of China (Project No. 51908204).